\documentclass[a4paper,11pt]{article}
\usepackage{pos}

\newcommand{\eV}{\mathrm{\,eV}}

\newcommand{\osc}{\mathrm{osc}}

\renewcommand{\eqref}[1]{eq.~(\ref{#1})}

\newcommand{\Figref}[1]{Figure~\ref{#1}}

\newcommand{\beq}{\begin{equation}}
\newcommand{\eeq}{\end{equation}}

\title{Gravitational signatures of ALP dark matter fragmentation}

\author[a,b]{Aleksandr Chatrchyan}
\author[a,c]{Cem Eröncel}
\author*[a,d]{Matthias Koschnitzke}
\author[a,d]{Géraldine Servant}

\affiliation[a]{Deutsches Elektronen-Synchrotron DESY, Notkestr. 85, 22607 Hamburg, Germany}
\affiliation[c]{Nordita, Stockholm University and KTH Royal Institute of Technology, Hannes Alfvéns v\"ag 12, SE-106 91 Stockholm, Sweden}
\affiliation[c]{Istanbul Technical University, Department of Physics, 34469 Maslak, Istanbul, Turkey}
\affiliation[d]{II. Institut für Theoretische Physik, Universit\"{a}t Hamburg, Luruper Chaussee 149, 22761 Hamburg, Germany}

\emailAdd{aleksandr.chatrchyan@su.se}
\emailAdd{cem.eroncel@itu.edu.tr}
\emailAdd{matthias.koschnitzke@desy.de}
\emailAdd{geraldine.servant@desy.de}

\abstract{The misalignment mechanism for axion-like particles (ALPs) is a leading explanation for dark matter. In this work we investigate ALPs with non-periodic potentials, which allow for large misalignment of the field from the minimum and make it possible for ALPs to match the relic density of dark matter in a large part of the parameter space. Such potentials give rise to self-interactions which can trigger an exponential growth of fluctuations in the ALP field via parametric resonance, leading to the fragmentation of the field. The fluctuations later collapse to halos that can be dense enough to produce observable gravitational effects. These effects would provide a probe of dark matter even if it does not couple to the Standard Model (or too feebly). We determine the relevant regions of parameter space in the (ALP mass, decay constant)-plane and compare predictions in different axion fragmentation models. These proceedings are a short version of  \cite{Chatrchyan:2023cmz}.}

\FullConference{The European Physical Society Conference on High Energy Physics (EPS-HEP2023)\\
 21-25 August 2023\\
Hamburg, Germany\\}

\begin{document}
\begin{flushright}
	DESY-24-024 \\
\end{flushright}
\maketitle

\section{Fragmented Dark Matter from ALPs with Non-Periodic Potentials}

ALPs typically arise as the angular degree of freedom of a complex field with a broken global $U(1)$-symmetry or as
the lowest-level state of a higher-dimensional gauge field in string theory compactifications. The potential arises from non-perturbative effects (for example by QCD-instantons for the QCD-axion \cite{Peccei:1977hh}). If the field is displaced from the minimum in the early universe, it remains frozen for $H>m_a$, where $m_a$ is the axion mass and $H$ the Hubble parameter, and behave as dark matter (DM) once the oscillations begin. Fragmentation, which refers to the effective transfer of energy from the zero-mode into high momentum modes \cite{Fonseca:2019ypl_fragmentation}, is possible in the cosine potential for initial field values tuned close to the top. If the axion has initial kinetic energy, it can cross multiple potential barriers and undergo fragmentation.

\begin{figure}[h!]
    \centering
    \includegraphics[width=0.8\textwidth]{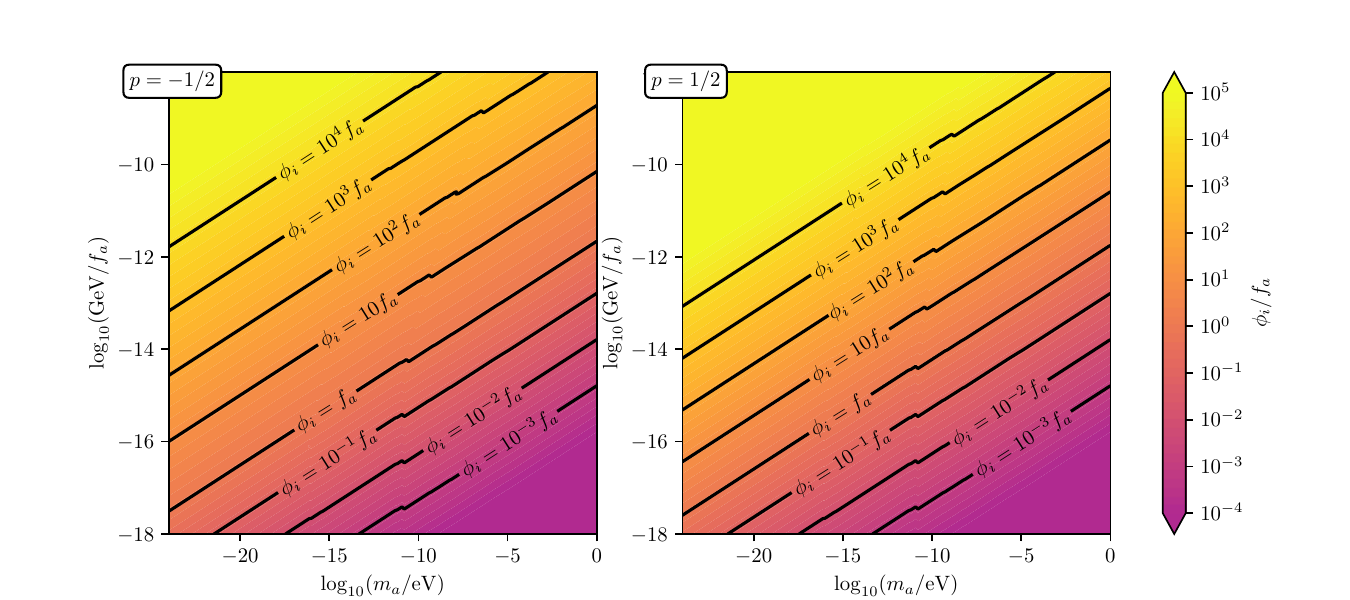}
    \caption{The initial field value $\phi_i/f_a$ required to match the dark matter relic density at each point in the $(m_a,f_a)$-plane for ALPs with the non-periodic potential. We set $p=-1/2$ and $p=1/2$ in the left and the right panels, respectively. Figure taken from \cite{Chatrchyan:2023cmz}.}
    \label{fig:monodromy_relic_plots}
\end{figure}

ALPs with non-periodic potentials are motivated by axion-monodromy in string theory \cite{Silverstein:2008sg,Dong:2010in_Monodromy3, PhysRevD.82.046003_Monodromy1} and
by ALPs coupled to pure Yang-Mills gauge fields in the large $N$ limit \cite{Witten:1980sp_Large_N_1,Witten:1997sc}. In these models, the full potential is periodic, but multi-branched, where different branches correspond to different configurations of the gauge fields. We assume that transitions between the branches are suppressed and the ALP $\phi$ effectively lives in a single, non-periodic branch. We take as effective potential 
\begin{equation}\label{eq:monodromy_potential}
    V(\phi)=\frac{m_a^2f_a^2}{2p}\left[\left(1+\left(\frac{\phi}{f_a}\right)^2\right)^p-1\right],
\end{equation}
with constant mass $m_a$ and decay constant $f_a$, choose $p=-1/2$ or $p=1/2$. Equations of motion for the homogeneous field $\phi$ and linear perturbations $u_k$ are: 
\begin{equation}
    \ddot{\phi}(t)+3H\dot{\phi}(t)+V'(\phi)=0\,\quad\mathrm{and}\quad
    \ddot{u}_k(t)+3H\dot{u}_k(t)+\left(\frac{\vec{k}^2}{a^2}+V''(\phi)\right)u_k(t)=0,
\end{equation}
where $a$ is the scale factor of the universe. 
The onset of oscillations, with respect to the standard misalignment mechanism in a periodic potential, is delayed for large initial field values $\theta_i\equiv\phi_i/f_a>1$, which allows the ALPs to match the DM relic density, $\Omega_{\mathrm{DM}}h^2\approx0.12$, in a large part of the $(m_a,f_a)$-parameter space, see \Figref{fig:monodromy_relic_plots}.

The oscillating homogeneous field can drive parametric resonance in the fluctuations. Since it is important to know when the linear approximation fails and we need to rely on a non-linear analysis, we define the \textit{critical field value} $\phi_{\mathrm{crit}}$  at which $\rho_{\mathrm{fluc}}/\rho_0$ becomes larger than unity for all initial field values $\phi_i>\phi_{\mathrm{crit}}$. In other words, the critical value is the smallest initial field value at which fragmentation happens. We find $\phi_\mathrm{crit}\approx 3f_a$ for $p=-1/2$ and $\approx 8f_a$ for $p=1/2$.

\section{Power Spectrum}

\begin{figure}[h!]
    \centering
    \includegraphics[width=0.5\textwidth]{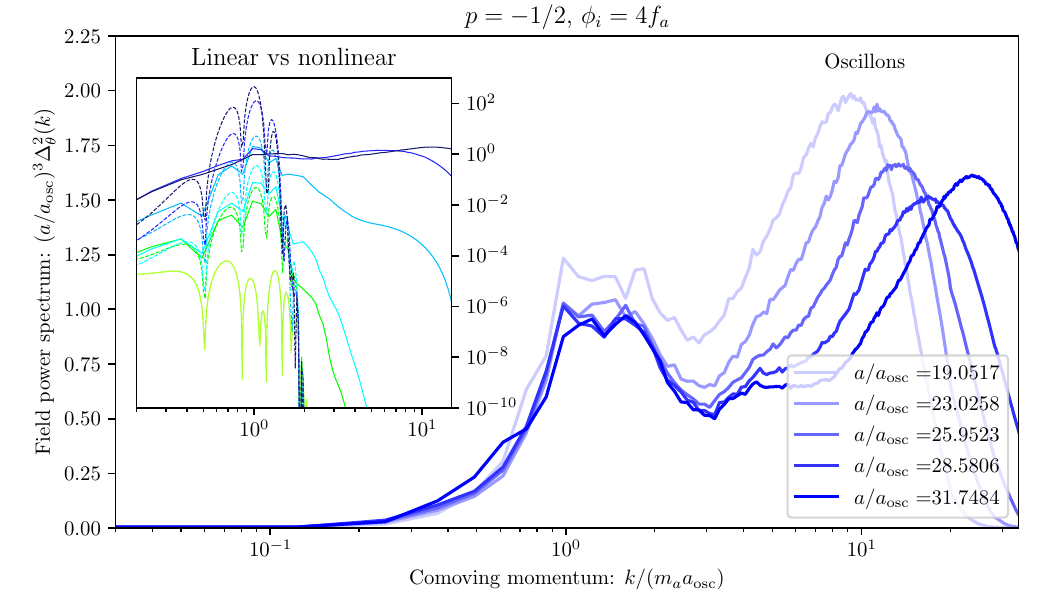}
    \caption{Snapshots of the dimensionless field power spectrum  $\Delta^2_{\theta}(t,k)$ at different $a$ for $\theta_i=4$ and $p=-1/2$, demonstrating the oscillon peak at high momenta and the fluctuation peak at small momenta. The inset compares the linear (dashed lines) and the lattice (solid lines) evolution at scale factors $a/a_{\osc}$ between $4$ and $24$, illustrating how non-linear effects slow down and spread out the growth of low-momentum fluctuations. Figure taken from \cite{Chatrchyan:2023cmz}.}
    \label{fig:field_power_spectrum2}
\end{figure}

If $\theta_i=\phi_i/f_a >1$, the anharmonicity of the potential leads to parametric resonance, which can transfer all the energy to higher $k$-modes. We employ a fully non-linear lattice simulation, which shows the formation of oscillons for large $\theta_i$. The field power spectrum separates into contributions from oscillons and the unbound fluctuations, see \Figref{fig:field_power_spectrum2}. Since oscillons decay at a later stage, we remove their contribution to the energy density power spectrum, which is defined as:
\begin{equation}
    \quad\Delta_{\delta}^2(k)=\frac{k^3}{2\pi^2}P(k)=\frac{k^3}{2\pi^2}\frac{<\left|\rho_k\right|^2>}{\mathcal{V}{\rho}^2},
\end{equation}
where $\rho_k$ is the Fourier transform of the energy density $\rho$ and $\mathcal{V}$ is a volume factor that drops out in the end.

\section{Halo Mass Function and Halo Spectrum}

\begin{figure}[h!]
	\centering
	\includegraphics[width=0.37\textwidth]{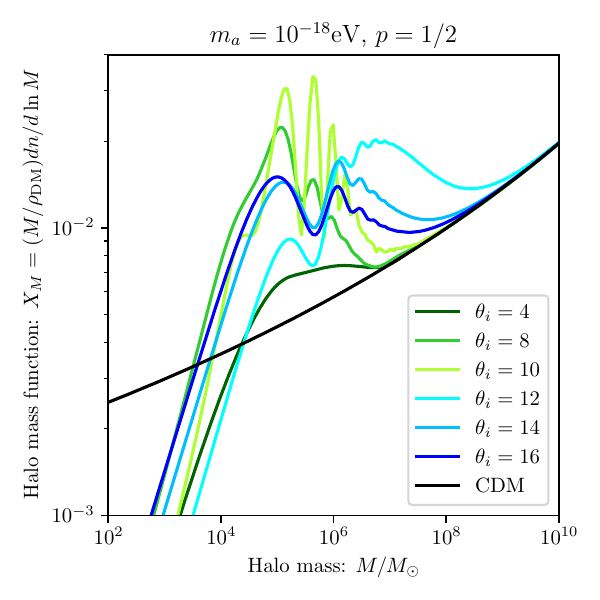}
 	\includegraphics[width=0.37\textwidth]{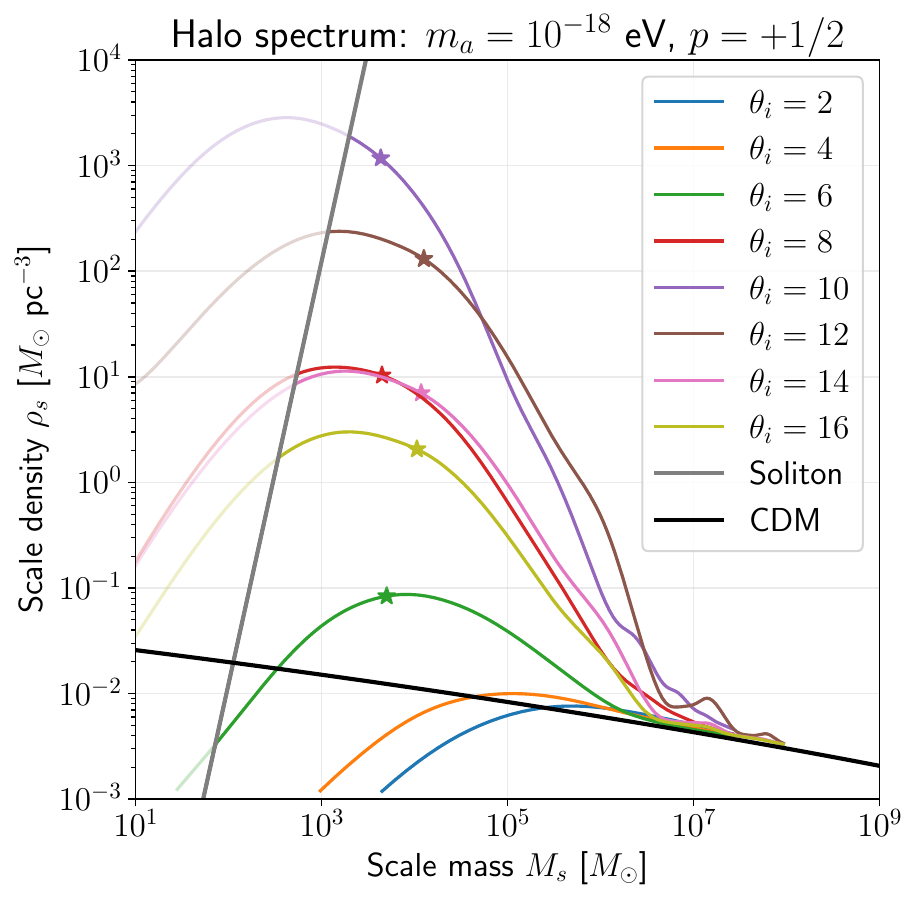}
	\caption{Left: The halo mass function $X_M = (M/\rho_{\mathrm{DM}}) d n/(d \ln M)$.  Right: Halo spectra, showing the scale density $\rho_s$ as function of the scale mass $M_s$. The soliton line for this mass is shown as straight grey line. In both plots we set $m_a=10^{-18}\eV$, use \eqref{eq:monodromy_potential} with $p=1/2$ and compare different initial field values $\theta_i = \phi_i/f_a$. Also shown are the CDM-lines in black. Figures taken from \cite{Chatrchyan:2023cmz}.}
	\label{fig:HMF}
\end{figure}

With the Press-Schechter (PS) formalism \cite{Press:1973iz}, we can find the abundance of halos per halo mass $M$. The power spectrum is used to find the variance $\sigma_R^2$ of the smoothed density contrast $\delta_R(\vec{x})=
\int d^3x' W_R(\vec{x}-\vec{x'})\delta(\vec{x'})$ on length scale $R$ by using an appropriate window function $\tilde{W}_R(\vec{k})$: 
\begin{equation}
     \sigma_R^2\equiv\frac{1}{\mathcal{V}}\int d^3x <\delta_R(\vec{x})^2> = \frac{1}{2\pi^2}\int dk\,k^2 |\tilde{W}_R(\vec{k})|^2P(k).
\end{equation}
From this we find the halo mass function: 
\begin{equation}
    \frac{dn}{dM}=-\frac{\rho_0}{M}\sqrt\frac{2}{\pi}\frac{\delta_c}{\sigma_R^2}\frac{d\sigma_R}{dM}\exp\left(\frac{-\delta_c^2}{2\sigma_R^2}\right),
\end{equation}
where $\rho_0$ is the DM density and $\delta_c$ is the critical overdensity for collapse.
\Figref{fig:HMF} shows this for different initial field values $\theta_i$ and compares it to the halo mass function of cold dark matter (CDM). 

The spherical collapse model \cite{Kolb:1994fi} and the PS formalism make it possible to find the halo spectrum, giving the scale mass $M_S$ and scale density $\rho_s$ of ALP halos collapsed to NFW-halos \cite{Navarro:1995iw}. We show examples in \Figref{fig:HMF}. Lighter axions form more massive halos, due to their larger Compton wavelength.
A peak in the halo spectra arises, because at smaller scales, we expect the formation of solitons due to gradient pressure, which induces a cutoff on the maximum density predicted for NFW-halos.

\section{Observational Prospects}

\begin{figure}[h!]
    \centering
    \includegraphics[width=0.8\textwidth]{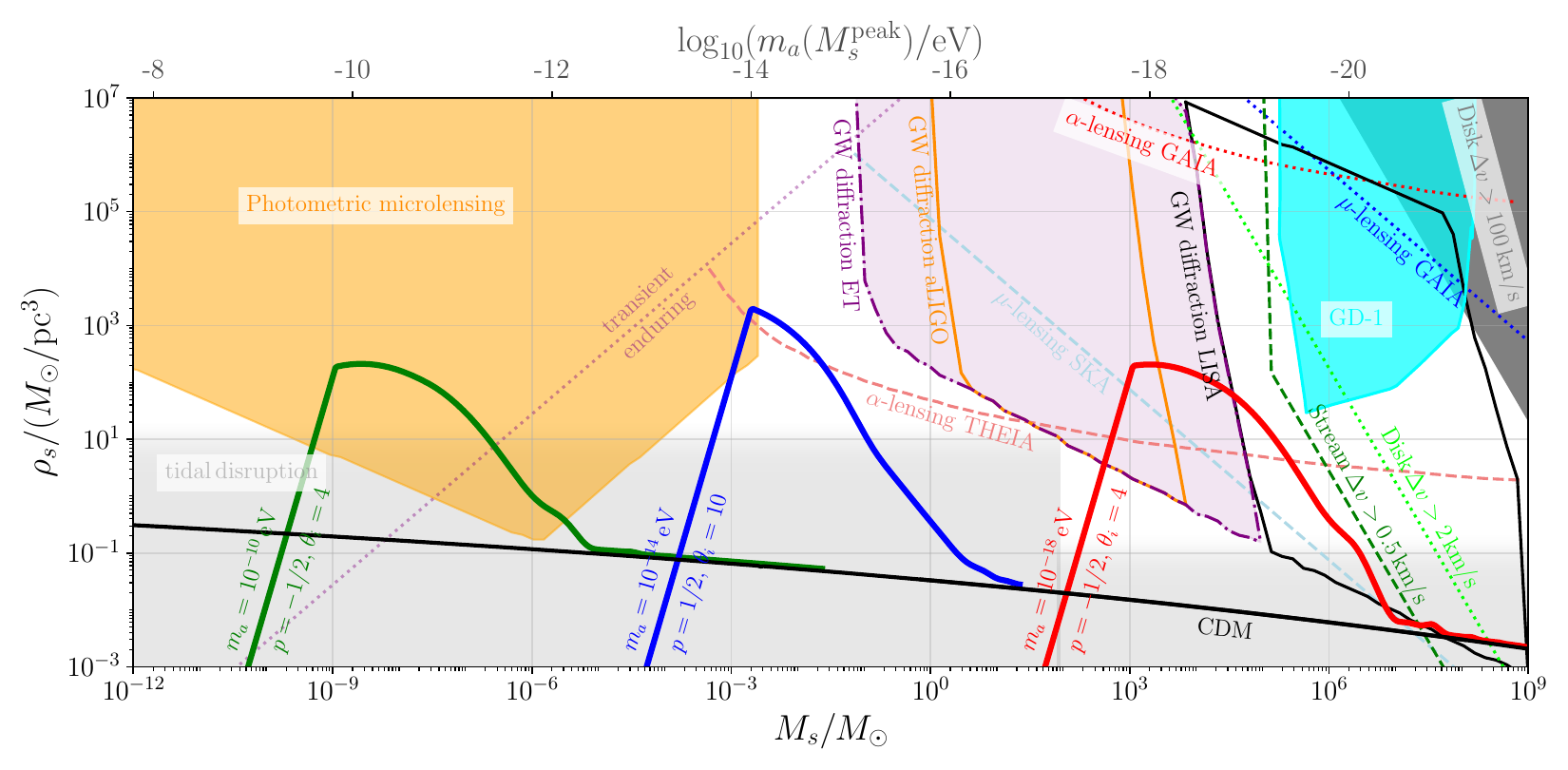}
    \caption{The thick lines show three of the halo spectra with the densest characteristic halos, with parameters specified on the lines. The CDM line is given in black. The grey region shows for which $M_S$ and $\rho_s$ halos are likely tidally disrupted within the Galaxy. The other coloured regions and lines depict regions in which we could see gravitational effects of characteristic halos with given $M_S$ and $\rho_s$ if they made up $30\%$ of all DM. See \cite{Chatrchyan:2023cmz}, from which this figure is taken, for details.}
    \label{fig:observational_prospects_03}
\end{figure}

The halos reach densities that can be tested by several purely gravitational observational prospects shown in \Figref{fig:observational_prospects_03}: 
Photometric microlensing are irregularities in the light curve of highly magnified stars that transit a caustic curve. Diffraction of gravitational waves (GWs) are distortions in the GW signals from black hole mergers caused by dark matter halos that could be observable by aLIGO, LISA or ET. Astrometric weak gravitational lensing are local effects in the angular velocity or acceleration of stars due to lensing effects possibly observable by GAIA, SKA or THEIA. Local gravitational perturbations refer to a significant change in the velocity of stars in stellar streams, the disk or the Galactic halo that ALP halos could cause.

\begin{figure}[h!]
    \centering
    \begin{minipage}{0.4\textwidth}
        \includegraphics[width=\textwidth]{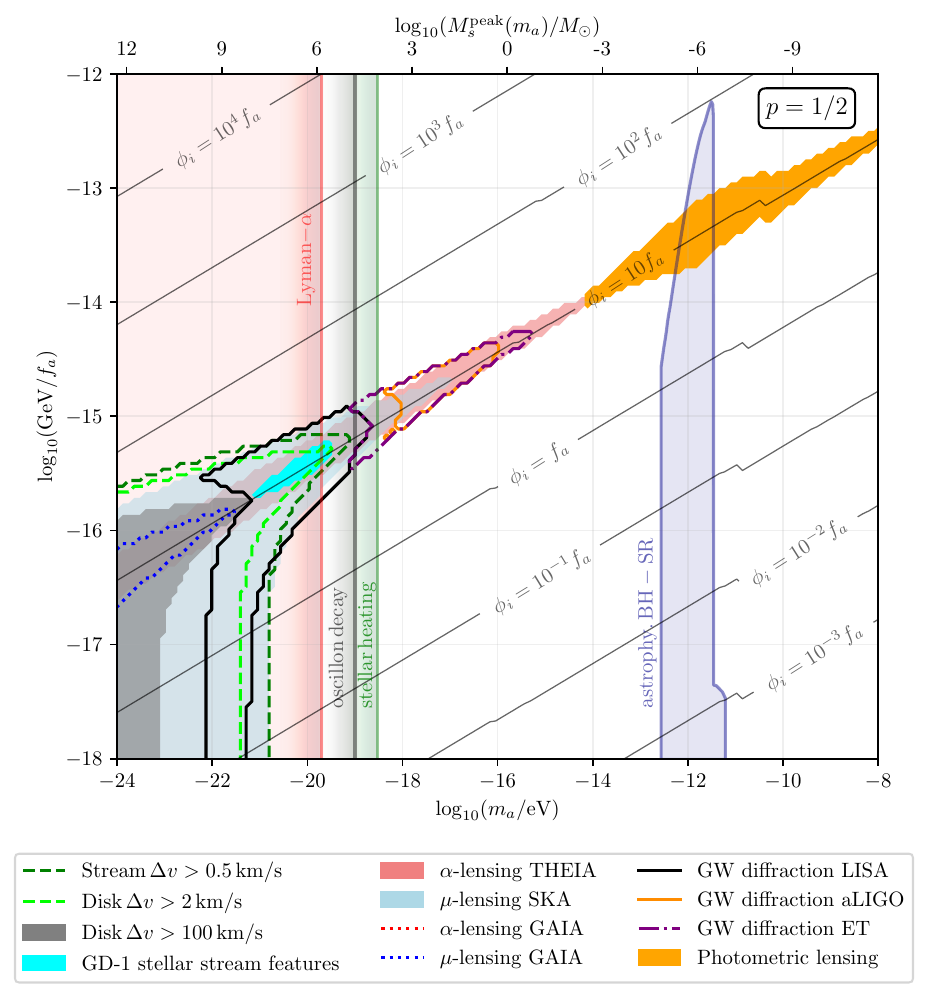}
    \end{minipage}
    \hfill
    \begin{minipage}{0.5\textwidth}
        \centering
        \includegraphics[width=\textwidth]{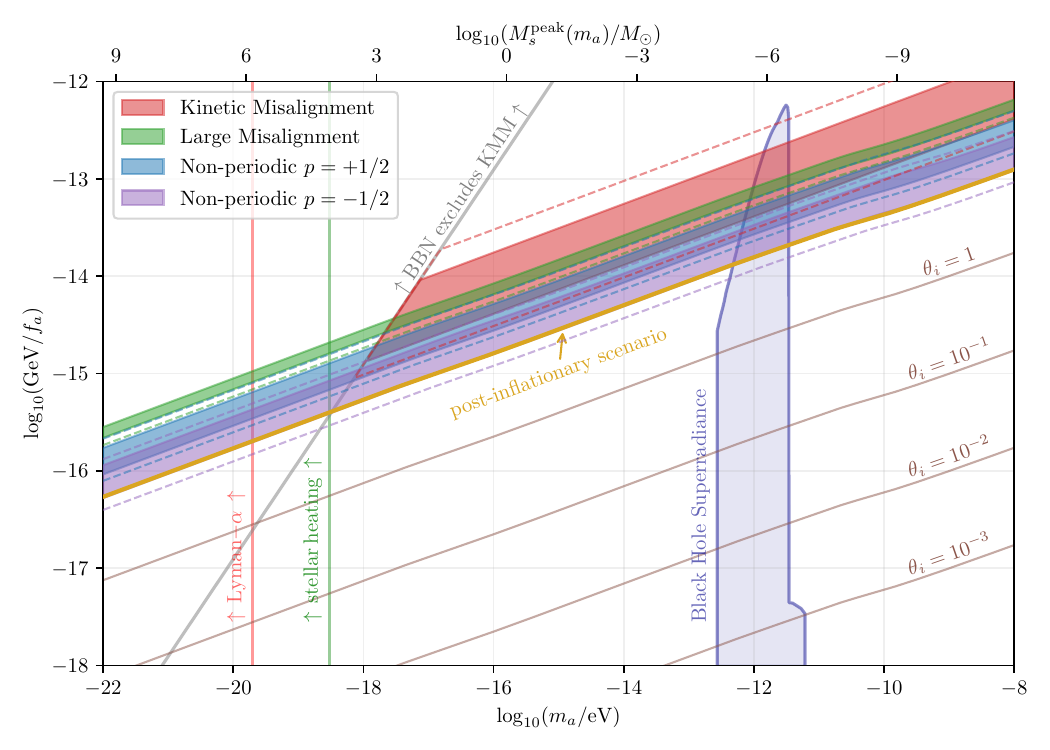}
    \end{minipage}
   
    \caption{Left: Regions in the $(m_a,f_a)$-plane in which ALP DM with potential \eqref{eq:monodromy_potential} with $p=1/2$ can lead to observable gravitational effects, assuming that 30\% of all of DM is inside of characteristic halos with masses within one decade around $M_s$. The corresponding initial field values $\theta_i$ are shown by the grey contour lines. The labelled regions in the plot show different constraints and the minimum mass at which oscillons decay before matter-radiation equality. Right: Regions in the ALP DM parameter space where the parametric resonance might create halos whose scale densities fulfill $\rho_s\gtrsim 10 M_{\odot}\,\textrm{pc}^{-3}$ (colored regions) or $\rho_s \gtrsim M_{\odot}\,\textrm{pc}^{-3}$ (dashed lines) in different scenarios. Figures taken from \cite{Chatrchyan:2023cmz}.}
    \label{fig:observations_m_f_plane_XM_0.3}
\end{figure}

When we assume that the ALP makes up all of DM, an extrapolation of the previous results to arbitrary $\theta_i$ allows us to project in which part of the $(m_a,f_a)$-plane the densest halos expected to form would lead to observational signatures, shown in \Figref{fig:observations_m_f_plane_XM_0.3}. We observe that dense halos form most efficiently when
the initial field value is close to the critical field value. For higher field values fluctuations are distributed among higher momentum modes, which collapse later due to the Jeans length and form less dense halos.  This means that observing effects of dense halos could \textit{give a handle to constrain axion mass and coupling}. We also compared the regions in the $(m_a,f_a)$-plane where dense halos are expected to form in other scenarios: fine-tuning in a cosine potential (Large Misalignment) \cite{Arvanitaki:2019rax}, the Kinetic Misalignment Mechanism \cite{Eroncel:2022efc,Eroncel:2022vjg} and post-inflationary models \cite{Enander:2017ogx}. These models predict dense halos in a similar parameter region, see \Figref{fig:observations_m_f_plane_XM_0.3}.

\section*{Acknowledgements}

\begin{footnotesize}
This work is supported by the Deutsche Forschungsgemeinschaft under Germany’s Excellence Strategy – EXC 2121 ,,Quantum Universe“ – 390833306. This work has been produced benefiting from the 2236 Co-Funded Brain Circulation Scheme2 (CoCirculation2)
of The Scientific and Technological Research Council of Turkey T\"UB\.{I}TAK (Project No: 121C404).

\bibliographystyle{JCAP}
\bibliography{references}


\end{footnotesize}
\end{document}